\documentclass[
showpacs,preprintnumbers,amsmath,amssymb]{revtex4}
\usepackage{amsmath}
\usepackage{float}
\usepackage{graphicx}
\usepackage[]{caption}
\usepackage{tensor}
\usepackage[toc,page]{appendix}
\usepackage{color,soul}

\begin{document}

\title{RADIATION FORCES AND THE ABRAHAM-MINKOWSKI PROBLEM}

\author {Iver Brevik\footnote{E-mail:iver.h.brevik@ntnu.no}}

\medskip

\affiliation{Department of Energy and Process Engineering, Norwegian University
of Science and Technology, N-7491 Trondheim, Norway}

 \today

\begin{abstract}

Recent years have witnessed a number of beautiful experiments in radiation optics. Our purpose with this mini-review is to highlight some developments of radiation pressure physics in general, and thereafter to focus on the importance of the mentioned experiments in regard to the classic Abraham-Minkowski problem. That means, what is  the "correct" expression for electromagnetic momentum density in continuous matter. In our opinion one sees relatively often that authors  over-interpret the importance of their experimental findings  with respect to the momentum problem. Most of these experiments are actually unable to discriminate between these energy-momentum tensors  at all, since they can be easily described in terms of force expressions that are common for Abraham and Minkowski.  Moreover, we emphasize the inherent ambiguity in applying formal conservation principles  to the radiation field in a dielectric, the reason being that the electromagnetic field in matter is only a subsystem which has to be supplemented by the mechanical subsystem to be closed.  Finally, we make some suggestions regarding the connection between macroscopic electrodynamics and the Casimir effect, suggesting that there is a limit for the magnitudes of cutoff parameters in QFT related to surface tension in ordinary hydromechanics.

\end{abstract}
 \maketitle

\section{Introduction}
\label{ChaptIntro}

The classic Abraham-Minkowski problem \cite{abraham09,abraham10,minkowski10} has in recent years attracted considerable interest. The core of the problem is to decide upon what is the most correct (or most convenient) expression for the electromagnetic momentum density in matter. This interest is quite understandable, as this concerns a fundamental aspect of the classical as well as the quantum mechanical theory of macroscopic electrodynamics, and also because the development of very accurate experimental techniques has made a deeper understanding of these phenomena more possible.

The purpose of the present mini-review is first to elucidate some aspects of radiation pressure physics in general, especially of its  recent developments. We here include also electrostriction effects, which appear in inhomogeneous regions when the compressibility of the medium is taken into account. The Abraham-Minkowski problem - an important arena for radiation pressure physics - is however usually restricted to cases where electrostriction is omitted. In those cases inhomogeneities  appear only in the dielectric boundary regions. We will in the Abraham-Minkowski  context consider various experiments - real ones as well as Gedanken ones - and try to give  a critical assessment of their importance for the present problem, in view of the circumstance that   some of the claims made in the literature are actually in conflict with each other.  Because of readability we  also wish  to cover some  basic  elements of  the electromagnetic theory in continuous media. This is made in the following section.

We may list some references, beginning deliberately with our own  1979 review \cite{brevik79}. A collection of newer  papers can be found in Refs.~\cite {pfeifer07,hinds09,barnett10,baxter10,barnett10a,mansuripur10,milonni10,kemp11,rikken11,griffiths12,testa13,mansuripur13,aanensen13,astrath14,leonhardt14,
brevikhoye14,barnett15,zhang15,kemp15,sheppard16,wang16,nesterenko16,nesterenko16a,brevik17,choi17,silveirinha17,partanen17,partanen17a,medina17}. We will return to some of these below.

In the first part of the paper, in Section III, we   focus on  electrostriction. As is known, this kind of force in a medium, usually a liquid, has often been overlooked (together with its magnetic counterpart), perhaps unjustifiably, because it plays an important  role for the local pressure in a liquid but usually does not contribute to the total force on a dielectric body if surrounded by a vacuum (air).

In Sections IV and V we briefly survey some of the conventional radiation force experiments and their relevance for the Abraham-Minkowski problem. There are actually not so many experiments that are able to give definite information about the electromagnetic momentum in matter.  We highlight some cases that we find important in this context. As already mentioned in many cases the experiments - although giving impressive results - are based upon properties that are common for the Abraham and Minkowski alternatives, and therefore not  connected with electromagnetic momentum. At the end of Sect.~V we present a proposal for how to extend the classic torque experiment of Walker {\it et al.} (1975) into the optical region.

In Section VI we  emphasize the limiting power of applying  conservation principles  to derive the "correctness" of one of the competing  momentum expressions. Arguments built upon this principle   are only able to  lead to  mathematical identities, not to a physical proof. The physical reason for this is that the electromagnetic field in a medium is a sub-system, which has to be supplemented with the material sub-system to form a closed system for which the conservation principles are more powerful. These arguments may appear trivial but we think they are justified nevertheless, since strong claims based upon symmetry principles are presented relatively often even in the contemporary  literature.

 In the final Conclusion section VII we sum up, and make  some remarks on a possible connection to the Casimir effect. There are arguments related to hydrodynamics that indicate that the cutoff parameters used in regularization procedures in quantum field theory (time-splitting) are restricted in magnitude. Actually, a simple numerical estimate shows that the elementary cutoff separation may be of the order of atomic dimensions. In our opinion this numerical coincidence to the order-of-magnitude level is worth attention.

\section{General electromagnetic force density expression, and  connection with the energy-momentum tensors}

Assume that the medium, in general compressive, is a homogeneous and isotropic fluid,  with density $\rho$. We take the constitutive relations in the form ${\bf D}=\varepsilon \varepsilon_0  {\bf E}, ~{\bf B}=\mu \mu_0 {\bf H}$, so that the permittivity $\varepsilon$ and the permeability $\mu$ become nondimensional quantities. We begin by writing down the expression for the basic electromagnetic force density $\bf f$,
\begin{align}
{\bf f}=&\rho{\bf E}+{\bf J\times B}+\frac{1}{2}\varepsilon_0 {\bf \nabla}\left[ E^2\rho_m\frac{\partial \varepsilon}{\partial \rho_m}\right]+
\frac{1}{2}\mu_0 {\bf \nabla}\left[ H^2\rho_m\frac{\partial \mu}{\partial \rho_m}\right] \notag \\
&-\frac{1}{2}\varepsilon_0 E^2{\bf \nabla} \varepsilon -\frac{1}{2}\mu_0H^2{\bf \nabla}\mu +\frac{n^2-1}{c^2}\frac{\partial}{\partial t}{\bf (E\times H)}. \label{1}
\end{align}
Here the first and second terms are standard expressions when $\rho$ is the charge density and $\bf J$ the current density. In connection with radiation problems, one can usually put $\rho=0,\, {\bf J=0}$.

 The third and fourth terms are the electrostriction and magnetostriction terms, respectively, where $\rho_m$ is the mass density of the fluid.   In Eq.~(\ref{1}) we have not specified under what conditions the partial derivatives are take; usually one means the isothermal derivative. If the fluid is nonpolar, one can simply make use of the Clausius-Mossotti equation to calculate the derivative. As mentioned above  we can usually omit the striction terms in practical applications, since being spatial  gradient terms they can be transformed into surface integrals giving zero total force   if the body is surrounded by a vacuum.

 The fifth and sixth terms in Eq.~(\ref{1}) are acting typically in the boundary regions of dielectric bodies.  In optical cases ($\varepsilon>1,\, \mu=1)$ these forces act in the direction of the optical thinner region, which means outwards from the body. The expressions hold for solid materials also. In the  case with anisotropic media, the expressions are easily generalized. These kind of forces have been measured experimentally in a number of cases.

 The  last term in Eq.~(\ref{1}) is the famous Abraham term, with $n^2=\varepsilon \mu$.  It is the central issue in the debate around the "correct" energy-momentum tensor in macroscopic electrodynamics. In contrast to the previous case, experimental verifications of the Abraham term are scarce. There exist actually many investigations of various kinds, both theoretical and experimental, claiming to have "solved" the Abraham-Minkowski problem.  Many of these statements in the literature are however incorrect, in the sense that the conclusions are stronger than justified by the formalism itself.

 \bigskip

 We will in the following consider some of the items above in more detail. Let us first relate, however,  the force expression (\ref{1}) to the two most well-known energy-momentum alternatives, viz. those of Minkowski and Abraham.  In a four-dimensional formulation, the four-force density has to be derived from the electromagnetic energy-momentum tensor $S_{\mu\nu}$ as
 \begin{equation}
 f_\mu = -\partial_\nu S_{\mu\nu}, \label{2}
 \end{equation}
 the indices $\mu$ and $\nu$ running from 1 to 4 (we use the same conventions as in Ref.~\cite{brevik79} or in M{\o}ller's book \cite{moller72}). The Minkowski and Abraham cases correspond to $S_{\mu\nu}=S_{\mu\nu}^{\rm M}$ and   $S_{\mu\nu}=S_{\mu\nu}^{\rm A}$.

 Consider first $ S_{\mu\nu}^{\rm M}$. Its spatial components are
 \begin{equation}
 S_{ik}^{\rm M}= -E_iD_k-H_iB_k+\frac{1}{2}\delta_{ik}({\bf E\cdot D+H\cdot B}), \label{3}
 \end{equation}
 while the energy flux density (Poynting vector) and momentum density are
 \begin{equation}
 {\bf S}^{\rm M}={\bf E\times H}, \quad {\bf g}^{\rm M}={\bf D\times B}. \label{4}
 \end{equation}
The energy density is
\begin{equation}
w^{\rm M}=\frac{1}{2}{\bf (E\cdot D+H\cdot B)}. \label{5}
\end{equation}
The Minkowski force density thus becomes
\begin{equation}
{\bf f}^{\rm M}= -\frac{1}{2}\varepsilon_0 E^2{\bf \nabla} \varepsilon -\frac{1}{2}\mu_0H^2{\bf \nabla}\mu, \label{6}
\end{equation}
showing that this force acts in the inhomogeneous regions only. In particular, the striction  forces are not accounted for in the Minkowski theory.

A definite advantage of the simple Minkowski theory is that it fits so nicely within the formalism of canonical field theory. It is thus a most natural starting point for the construction of quantum electrodynamics in media. This point is further discussed, for instance, in Ref.~\cite{brevik17}.

In the Abraham case we have $S_{ik}^{\rm A}=S_{ik}^{\rm M},\, {\bf S}^{\rm A}={\bf S}^{\rm M}={\bf E\times H}, \, w^{\rm A}=w^{\rm M}$. The difference from the foregoing case occurs in the force density, which becomes
\begin{equation}
{\bf f}^{\rm A}= {\bf f}^{\rm M} +\frac{n^2-1}{c^2}\frac{\partial}{\partial t}{\bf (E\times H)}\mathrm{}. \label{7}
\end{equation}
That is, the core of the Abraham-Minkowski problem lies in the Abraham term, which is a volume term.  The other forces acting in the boundary regions are  without importance in this context.  The Abraham term is usually quite small, rapidly fluctuating, and difficult to detect experimentally. As in the Minkowski case, the Abraham theory is unable to encompass the striction forces.

 Note that the force expression (\ref{1})   is  effectively the Abraham force augmented with the striction  forces. It is also called the Helmholtz force. The expression is also in moreover  agreement with Landau and Lifshitz \cite{landau84}.

A  point that we want to highlight in this context  is the impressive numerical calculation of Partanen {\it et al.} \cite{partanen17}, about  the forces acting on an electromagnetic pulse propagating through a  liquid. The various contributions to the force (\ref{1}) were treated explicitly, and the small displacements of the particles were followed numerically.  In our opinion this is an important progress obtained in the field in the recent years.

\section{ Remarks on the electrostriction force}

 As already mentioned, the electrostrictive pressure term in a fluid has received only moderate attention in practice, due to its gradient form usually leading to a zero total force on a dielectric specimen. The striction is a {\it local} effect. In the field of optofluidics, its influence is however expected to increase. Imagine, as an example, the following situation: let a dielectric sphere be exposed instantaneously to a radiation beam at time $t=0$. The electrostriction force due to the beam can be written as $\bf \nabla \chi$, where
 \begin{equation}
 \chi= \frac{1}{2}\varepsilon_0E^2 \rho_m \frac{dn^2}{d\rho_m }=\frac{1}{6}\varepsilon_0E^2(n^2-1)(n^2+2) \label{8}
 \end{equation}
  may be called the electrostriction potential (we have here made use of the Clausius-Mossoti relation according to which the permittivity depends on the density only). Assume for simplicity that the force from the beam is the same all over the spherical surface. The occurs thus for $t>0$ a {\it compressive} force at the boundary so that a spherical sound wave moves inwards and creates a pressure maximum at the center, until it bounces off.  Imagine now that there is a medical substance  situated in a small sphere at the origin, protected weakly by thin glass walls. If the compressive wave is strong enough it may break the walls, thus permitting the substance to be delivered to the body at the desired location. (This only as a suggestion about making use of electrostriction in medical technology - we are not aware of this idea being used in practice.)

  As for earlier treatises on electrostriction, we mention a few papers \cite{brevik79,lai89,zimmerli99,ellingsen11}. Here we notice that the interesting paper of Zimmerli {\it et al.} \cite{zimmerli99}, dealing with electrostriction in a fluid under extraterrestrial conditions, is actually a generalization of the classic electrostriction experiment of Hakim and Higham \cite{hakim62} measuring the local increase of refractive index in a fluid exposed to a strong electric field (cf. also the discussion in Ref.~\cite{brevik79} on this point).

  To make our discussion more concrete, let us consider for simplicity a cylindrical symmetric situation in which a laser beam propagates in the $z$ direction through a homogeneous liquid. The power $P$ is taken to be switched on suddenly at $t=0$, and is thereafter held constant. Adopting cylindrical coordinates $(R, \theta, z)$ with the center of the beam at the origin, we have for a Gaussian beam the field distribution
  \begin{equation}
  E^2(R,z)= \frac{2P}{\pi \varepsilon_0 nc w^2(z)}\exp \left[ -\frac{2R^2}{w^2(z)}\right], \label{9}
  \end{equation}
  where $w(z)=w_0[ 1+z^2/l_R^2]^{1/2}$ is the beam radius. Here $c$ is the velocity of light in vacuum and  $l_R=\pi w_0^2/\lambda$, with $\lambda$ the wavelength in  the liquid, is called the Rayleigh length. We will assume, as usual in trapping experiments, that the waist $w_0$ amounts to  a few micrometers but is yet much larger than $\lambda$  so that $w(z)$ does not deviate very much from $w_0$.

  From Eqs.~(\ref{8}) and (\ref{9}) we have for the electrostriction potential at the center of the beam $R=0,\, z=0$,
  \begin{equation}
  \chi(0)= \frac{(n^2-1)(n^2+2)P}{3\pi ncw_0^2}. \label{10}
  \end{equation}
  The linearized Euler equation describing the electrostriction pressure wave is
\begin{equation}
\rho_m\partial_t{\bf v}=-{\bf \nabla}p+{\bf \nabla \chi}. \label{11}
\end{equation}
As $\rho_m$ varies only slightly, we get from this equation ${\bf \nabla \times v}=0$ permitting us to introduce a velocity potential $\Phi$, whereby ${\bf v}={\bf \nabla} \Phi$. Thus $p=-\rho_m\partial_t\Phi+\chi$, which can be processed further using the continuity equation for sound,
\begin{equation}
\partial_t\rho_m+{\bf \nabla \cdot}(\rho_m{\bf v})=0. \label{12}
\end{equation}
With $u= [(\partial  p/\partial \rho_m)_S]^{1/2}$ denoting the velocity of sound, we thus obtain the governing equation for $\Phi$,
\begin{equation}
\nabla^2\Phi-\frac{1}{u^2}\frac{\partial^2 \Phi}{\partial t^2}=-\frac{1}{\rho u^2}\frac{\partial \chi}{\partial t}. \label{13}
\end{equation}
As the beam is suddenly switched on at $t=0$, we have $E^2({\bf r},t)=E^2({\bf r}) \theta(t)$, so with $d\theta/dt=\delta(t)$ we can write the time-dependent velocity potential at the origin as
\begin{equation}
\Phi(0,t)=\frac{1}{4\pi \rho_m u^2}\int \frac{\chi({\bf r}^\prime)}{r^\prime}\delta \left( t-\frac{r^\prime}{u}\right) d^3r^\prime, \label{14}
\end{equation}
where $r^\prime=(R^2+z^2)^{1/2}$ is the magnitude of the vector ${\bf r}^\prime$ to the source point. Inserting the above expressions for $\chi$ and $l_R$ we then have
\begin{equation}
\Phi(0,t)=\frac{(n^2-1)(n^2+2)P}{3\pi \rho_m ucw_0^2}\int_0^{ut} \frac{dz}{1+z^2/l_R^2}\exp \left[-\frac{2(u^2t^2-z^2)}{w_0^2(1+z^2/l_R^2)}\right], \label{15}
\end{equation}
leading to the following pressure at the origin
\begin{equation}
p(0,t)=\chi(0)\left\{ \frac{u^2t^2}{l_R^2+u^2t^2}+\frac{4ut}{w_0^2}\int_0^{ut}\frac{dz}{(1+z^2/l_R^2)^2}\exp \left[ -\frac{2(u^2t^2-z^2)}{w_0^2(1+z^2/l_R^2)} \right]\right\}. \label{16}
\end{equation}
The above formalism contains essentially extracts from Refs.~\cite{brevik79} and \cite{ellingsen11}. Numerical calculations of the expression (\ref{16}) show that $p(0,t)$ approaches $\chi(0)$ when $t\rightarrow \infty$.

The time scale is here actually quite small, as it is determined by the time $\tau$ that sound needs to traverse the cross section of the beam. If we assume a typical waist radius of $w_0 = 4.5~\mu$m, we see that $\tau \sim 7~$ns. Calculated curves for $p(0,t)$ show that the maximum pressure occurs at about half of this, when $t\sim 3~$ns, and that after about 10 ns the electrostrictive wave dies out. Thus the compensating elastic pressure in the liquid has ample time to build itself up long before  physical effects such as  surface deformations of a dielectric cylinder or a sphere become visible. The electrostrictive, inward-directed pressure wave is obviously a transient effect. Once pressure equilibrium is established, the presence of the  electrostriction force    is usually unobservable in radiation pressure experiments. This is the reason why electrostriction is usually simply omitted.

From Fig.~9 in Ref.~\cite{brevik79} it follows that $p(0)_{\rm max} \approx 1.3 \times \chi(0)$ when $t\approx 3~$ns. The  excess pressure peak is thus about 30\%. Although we have for simplicity assumed that the  beam is stationary for $t>0$, it is reasonable to assume that pulsed lasers, with a high power $P$ but yet below the threshold  for nonlinear effects, would be actual here.  Let us follow Ashkin and Dziedzic in their classic paper \cite{ashkin73} in assuming $P=3~$kW. With $w_0=4.5~\mu$m as before, and with $n=1.33$ (water) we find
\begin{equation}
\chi(0)=1.14\times 10^5~{\rm Pa}, \label{17}
\end{equation}
which is close to one atmosphere. Thus
\begin{equation}
p(0)_{\rm max} \approx 3.4\times 10^5~{\rm Pa}. \label{18}
\end{equation}
A few times the atmospheric pressure is thus the excess electrostrictive pressure we may expect in practice.

\section{Radiation pressure experiments}

There exist by now a considerable amount of radiation pressure experiments in optics. It is probably right to say that the modern development began with the seminal work of Ashkin and Dziedzic \cite{ashkin73}, operating in water. They used a transverse mode frequency doubled  Nd:YAG laser at a free space  wavelength of $0.53~\mu$m, with peak power $3~$kW as mentioned above.  The duration of each pulse was about  60 ns. The waist $w_0$ was most likely about $4~\mu$m (a little larger than actually reported in their paper). The observed height of the outward bulge of the free surface was  $0.9~\mu$m. This  experiment  is discussed in detail also in Ref.~\cite{brevik79}.

This surface elevation was after all quite small, due to the high surface tension.  In order to enhance the effect, one may use a micellar liquid in the vicinity of the critical point. In that way one can reduce the surface tension by a factor of about one million, and the elevation accordingly can become much greater. This was the central element in the experiments reported  in Refs.~\cite{casner03,wunenburger11}. The elevation could be up to about $70~\mu$m at maximum laser power, which was about 1 watt. There are various papers discussing these experiments theoretically; cf. for instance Refs.~{\cite{aanensen13,hallanger05,birkeland08}.

The recent beautiful experiment of Astrath {\it et al.} \cite{astrath14} ought also to be highlighted. In principle, the method was the same as before, namely to observe the elevation of a water surface when illuminated by a vertical laser beam. A maximum of $30~$nm was observed (laser power $7~$W) and illustrated in various diagrams. A related paper is that of Capeloto {\it et al.} \cite{capeloto16}.

A special variant of radiation pressure experiments occurs if  the laser beam is directed towards a curved surface. The experiment of Zhang and Chang \cite{zhang88} is of this sort. Using pulsed lasers of energies 100 mJ they measured the deflection of water droplet surfaces.  Because of the lens effect the behavior of the front end and rear ends of the droplet was found to be quite different: a relative displacement of 2\% occurred at the front and up to 30\% at the rear. Theoretical studies of this experiment can be found in Refs.~\cite{lai89,brevik99,ellingsen13}.

Related to this is the experiment of Zhang {\it et al.} \cite{zhang15}, which analyzed  the deflection of mineral oil, as well as water,  surfaces when illuminated by a laser. The experiment checked whether the free surface formed a convex defocusing surface, or a concave focusing one. The conclusion of the experimentalists was that the Abraham momentum pressure is the correct one.

Another example of a recent radiation force experiment is that  of Choi {\it et al.} \cite{choi17}, which made  use of an adiabatic liquid-core optical fiber waveguide. It showed an unconventional way of measuring the force that a laser field exerts on a dielectric liquid surface:  the incident laser field induced a linear axial displacement of the air-liquid interface inside the hollow fiber, and this displacement was easily detectable. Also on that case it was claimed, as a result of the analysis,  that the Abraham expression is the correct one.

In our context the  important question is however: is it really true that  these radiation pressure experiments are able to discriminate between the Abraham and Minkowski momentum densities? In our opinion the answer is no, is spite of the arguments to the opposite just mentioned. Let us  go back to the basic force expression (\ref{1}): the electrostriction force, as we have seen, is a transient force that seeks to compress the liquid immediately after the imposition of the laser field. After a short time scale determined by the velocity of sound the counteracting elastic pressure will have had sufficient time to build itself up, and electrostriction will at stationary conditions at later instants not have any influence upon   observable effects at all. It is to be noted that radiation-induced  deformations  of surfaces are  in this context  slow processes.

 Moreover, the Abraham term in Eq.~(\ref{1}) simply fluctuates out at the high  optical frequencies. What is left in the force expression, is simply the term describing the optical force in regions where the refractive index varies, that means in practice the dielectric boundaries. This force is {\it common} for the Abraham and Minkowski tensors, and may conveniently be called the Abraham-Minkowski force ${\bf f}^{\rm AM}$,
\begin{equation}
{\bf f}^{\rm AM}= -\frac{1}{2}\varepsilon_0 E^2{\bf \nabla} n^2. \label{17}
\end{equation}
We have here assumed  usual conditions where $\rho$ and $\bf J$ are zero and where the material is nonmagnetic. The experiments referred to above must necessarily be describable in terms of the force (\ref{17}).

We mention finally the recent paper of  Kundu {\it et al}  \cite{kundu17}, which demonstrates how a horizontal graphene oxide sheet illuminated by a weak laser beam from above becomes  deflected downwards by the radiation pressure (power $P=1.4~$mW, deflection about 80 nm).  The outcome of the  experiment made the authors claim that the Abraham momentum is to be favored.  However, in our opinion this experiment belongs essentially to the same category as those considered above, and is describable in terms of the common Abraham-Minkowski  force acting at the boundaries, plus the additional Lorentz force acting in the interior due to the fact that graphene oxide has an appreciable conductivity. Its refractive index is complex, equal to  $\tilde{n}=2.4 +1.0i$ at wavelength 532 nm \cite{wang08}.    There is no direct connection with the electromagnetic momentum here, and the experiment is unable to discriminate between Abraham and Minkowski.

\section{On experiments testing the electromagnetic momentum}

From the above it should be clear that the electromagnetic momentum in matter is accessible to measurement only in special cases. In the following we will briefly cover three existing experiments, and in addition make one proposal. Our general picture is that it is the Minkowski momentum ${\bf g}^{\rm M}$ which is the total propagating momentum, being the sum of the field Abraham momentum ${\rm g}^{\rm A}$ and an accompanying mechanical momentum. The three first cases will show the existence of the Minkowski momentum.

\bigskip

\noindent 1. One prominent example is the accurate series of radiation pressure experiments made by Jones {\it et al} \cite{jones54,jones78} (cf. also the lucid exposition in Jones' book \cite{jones88}). What was measured was the pressure on a mirror immersed in a dielectric liquid, and this pressure was found to be proportional to the refractive index $n$ of the liquid. This is just as one would expect for the flowing Minkowski momentum (this experiment is discussed in more detail in Ref.~\cite{brevik79}).

\noindent 2. Our second example is the photon drag experiment of Gibson {\it et al.}  \cite{gibson80}. This experiment was made on a semiconductor, not a fluid. The main idea was that a longitudinal electric field can be produced by the momentum transfer from the radiation field to the electrons in the material, thereby causing charges to be driven down in the rod. The current in the rod must be zero under open circuit conditions, and so is a measurable voltage difference between the ends being in accordance with the prediction from the Minkowski momentum (cf. also the separate treatment of this experiment in Ref.~\cite{brevik86}).

\noindent 3.  Whereas the detection of the Minkowski momentum in two foregoing cases may be considered somewhat indirect, the radiation pressure experiment of Campbell {\it et al.} \cite{campbell05} provides a quite direct verification. Here the photon recoil momentum was observed in a Bose-Einstein condensate; this condensate having higher density than the atomic cloud cooled by a laser. The photon momentum was actually measured to be $\hbar k=\hbar n\omega/c$, precisely in accordance with the Minkowski prediction.

\noindent 4. One may ask: can the Abraham term, which we will call ${\bf f}^{\rm Aterm}$,
\begin{equation}
 {\bf f}^{\rm Aterm}=  \frac{n^2-1}{c^2}\frac{\partial}{\partial t}{\bf (E\times H)}, \label{18}
 \end{equation}
 be measured in optics? In a stationary field it can obviously not be detected since it is proportional to the time derivative of the Poynting vector and thus fluctuates out.  Consider, however,  the following proposal  (actually discussed earlier in Ref.~\cite{brevik10}), which    exploits that the  field energy flowing in microspheres, in the  so-called whispering gallery modes, is large.  The energy flux can surpass 100 W, close to the rim. Assume now that such a microsphere is suspended in the gravitational field and is fed with an intensity modulated laser beam, modulated at the same frequency $\omega_0$ as the mechanical oscillation frequency as the sphere. The system then becomes exposed to a vertical torque, caused by the Abraham force (\ref{18}).

 Numerical considerations using the values $P=100~$W, $\omega_0 =1000~$rad/s as input, indicate that the torque should be of order $10^{-19}~\rm{N~m}$. It is small, but of
 the same order as torques in micromechanics often discussed in the contemporary literature.

 This idea goes back to the classic experiments of Walker {\it et al.} \cite{walker75a,walker75b}, operating at  low frequencies for which the oscillations of a dielectric cylindrical shell of high permittivity (barium titanate, $\varepsilon \sim 3600$) suspended in the gravitational field were directly measurable. The medium in the shell was exposed to strong time-varying crossed electric and magnetic fields. The torque amplitude was of order $10^{-10}~$N~m. The idea we have sketched above is thus simply an optical analogy to that of the Walker {\it et al.} experiment.

\section{Remarks on the application of conservation principles}

The conservation principles in field theory, for energy, momentum, and center-of-mass velocity, are known to be powerful principles when applied to a physically  closed system. Quite often these principles have been applied to the radiation field in matter also - this idea seems to go back to the 1953 paper of Balasz \cite{balasz53} and is often called  Balasz gedanken experiments.  (Balasz himself argued for the correctness of Abraham instead of Minkowski by analyzing total momentum and center of mass.)  What we have to point out in this context, is that there is an ambiguity in these kinds of arguments, due to the fact that  the radiation field in matter is only a {\it subsystem}; it has to be supplemented by the mechanical subsystem in order to form a closed system.   In practice, the application of conservation principles to the electromagnetic subsystem will lead to an answer which is already fixed by physical assumptions implicitly brought into the formalism at an earlier stage. Effectively, one will usually end up with a mathematical identity.

Some recent papers dealing with Lagrangian methods and conservation principles can be found in Refs.~\cite{silveirinha17,dodin12,crenshaw14,crenshaw16}. We will not go into further details about  those topics here, but will show by an example how necessary it is to take care  when analyzing the conservation equations for a subsystem. In Ref.~\cite{silveirinha17}, arguing for the preference of the Abraham momentum, use was made of the relationship
\begin{equation}
{\bf g}={\bf S}/c^2, \label{19}
\end{equation}
which is often called the Planck principle of inertia of energy. Here $\bf g$ is the momentum density and $\bf S$ the energy flux density. This is an important relation, which is equivalent to the symmetry property of the fourth column and fourth line of the total energy-momentum tensor. But does the principle hold for the subsystems also? In general not so, as we find in the Abraham case
\begin{equation}
{\bf g}^{\rm A}= {\bf S}/c^2, \label{20}
\end{equation}
and in the Minkowski case
\begin{equation}
{\bf g}^{\rm M}=n^2{\bf S}/c^2, \label{21}
\end{equation}
where ${\bf S =E\times H}$. Now, in Eqs.~(16) and (19) in Ref.~\cite{silveirinha17} the Planck principle is chosen to hold for the mechanical subsystem. This means that the Planck principle is adopted for the radiation subsystem also, as the principle  necessarily has to hold for the total system. According to Eq.~(\ref{20}) the Abraham momentum is implicitly singled out in this way at an earlier stage of the argument, and must consequently turn up as the result of the formalism at the end.

There are several other examples from the literature that can be analyzed in the same manner. Our only point here has been  to emphasize the lack of unambiguity by this kind of reasoning.

\section{Conclusions. Relations to the Casimir effect. }

 The basic physical expression for the electromagnetic force density is in our opinion the expression (\ref{1}). The contributions  from the various component terms in the case of a traveling wave pulse have been strikingly shown in the numerical calculation of Partanen {\it et al.} \cite{partanen17}. This expression contains, in addition to the conventional force on charge densities $\rho$ and current densitie $\bf J$, the electrostrictive and magnetostrictive contributions, the force acting in regions of inhomogeneity (typically dielectric boundaries), and finally the Abraham term ${\bf f}^{\rm Aterm}$. The latter is given explicitly in Eq.~(\ref{18}), and is the core of the famous Abraham-Minkowski problem \cite{abraham09,abraham10,minkowski10}.

 Under normal circumstances, all the terms in Eq.~(\ref{1}) need not be taken into account. There are important special cases:

 \bigskip

 \noindent {\it The Abraham term omitted.} One can most often simply put ${\bf f}^{\rm Aterm}=0.$ In electrostatics or magnetostatics, this is evident. In radiation optics the same will be true, because the term fluctuates out under the high optical frequencies. Thus we obtain a reduced expression which is called the Helmholtz force, ${\bf f}^{\rm H}$,
 \begin{align}
{\bf f}^{\rm H}=&\rho{\bf E}+{\bf J\times B}+\frac{1}{2}\varepsilon_0 {\bf \nabla}\left[ E^2\rho_m\frac{\partial \varepsilon}{\partial \rho_m}\right]+
\frac{1}{2}\mu_0 {\bf \nabla}\left[ H^2\rho_m\frac{\partial \mu}{\partial \rho_m}\right] \notag \\
&-\frac{1}{2}\varepsilon_0 E^2{\bf \nabla} \varepsilon -\frac{1}{2}\mu_0H^2{\bf \nabla}\mu . \label{22}
\end{align}
Henceforth we put $\rho ={\bf J}=0$, and assume also that the medium is nonmagnetic, which is usually the case. The Helmholtz force reduces to
\begin{equation}
{\bf f}^{\rm H}=\frac{1}{2}\varepsilon_0 {\bf \nabla}\left[ E^2\rho_m\frac{\partial \varepsilon}{\partial \rho_m}\right]
-\frac{1}{2}\varepsilon_0 E^2{\bf \nabla} \varepsilon. \label{23}
\end{equation}
Here the electrostriction term requires special attention. In electrostatics, it was clearly measured in the Hakim-Higham experiment \cite{hakim62} as an excessive fluid pressure. In time-varying fields, one should note that the extension of the electrostriction formula is strictly speaking a hypothesis - the formula is after all derived rigorously for static fields only. We simply assume that it holds also for the time-dependent case also, in accordance with common practice. The need for experimental support has been pointed out earlier also \cite{nesterenko16a}.

 For optical frequencies, we are not aware of direct observations of the electrostriction force, although we have provided theoretical estimates of the transient excess pressure \cite{brevik79,ellingsen11}, especially applied to the Ashkin-Dziedzic radiation pressure experiment \cite{ashkin73}. An important point here is that the time scale is so low, of the order of nanoseconds when the linear transverse scale of the beam is of order micrometers. For larger times $t$ after the (assumed sudden) onset of the pulse at $t=0$ the electrostriction pressure meets a compensating elastic pressure and surface deflection effects become uninfluenced by the electrostriction.

\bigskip
\noindent {\it Electrostriction term omitted.} This is the classical Abraham-Minkowski problem, covered in Section II. For a nonmagnetic medium with $\rho={\bf J}=0$, the Minkowski force becomes
\begin{equation}
{\bf f}^{\rm M}= -\frac{1}{2}\varepsilon_0 E^2{\bf \nabla} \varepsilon,  \label{24}
\end{equation}
and the Abraham force
\begin{equation}
{\bf f}^{\rm A}=  -\frac{1}{2}\varepsilon_0 E^2{\bf \nabla} \varepsilon+\frac{n^2-1}{c^2}\frac{\partial}{\partial t}{\bf (E\times H)}\mathrm{}. \label{25}
\end{equation}
In most instances we can simply omit here the Abraham term - as mentioned it simply fluctuates out. Thus we are left with the expression for ${\bf f}^{\rm AM}$ given in Eq.~(\ref{17}) which, as mentioned, is common for the Abraham and Minkowski tensors.  Almost all of the modern radiation pressure experiments that we are aware of, can simply be described using that  expression. These experiments are thus not able to discriminate between Abraham and Minkowski.

We repeat, however, the idea presented at the end of Section  V using an intensity modulated laser beam propagating near the rim of a suspended microsphere, leading to a vertical torque.  If this idea could be realized experimentally, it would be the first case that the Abraham force is detected explicitly in optics.

\bigskip
\noindent {\it Properties of Minkowski's energy-momentum tensor.} As pointed out above, the Minkowski momentum ${\bf g}^{\rm M}$ in a wave is composed of the Abraham momentum ${\bf g}^{\rm A}$ and the accompanying mechanical momentum. The four-dimensional Minkowski energy-momentum tensor $S_{\mu\nu}^{\rm M}$ has three characteristic properties:

\noindent 1. For an electromagnetic field in a homogeneous medium it is divergence-free,
\begin{equation}
\partial_\nu S_{\mu\nu}^{\rm M}=0,  \label{25}
\end{equation}
meaning that the corresponding  total momentum components and the total energy form a four-vector \cite{moller72}. This makes it convenient to construct a relativistic covariant theory: in any inertial system one defines $S_{\mu\nu}^{\rm M}$ to be just the tensor that reduces to the known expression in the medium's rest frame.

\noindent 2. The second property is that the mentioned four-momentum is {\it spacelike}. It means, in a class of inertial systems the Minkowski field energy will be negative. A typical expression for this is seen in the Cherenkov effect: in the emitting particle's rest system the emitted energy is negative, in order to balance the positive particle energy obtained after the emission.

\noindent 3. The Minkowski tensor, per definition, has the same form in any inertial system.  The material velocity thus does not appear explicitly in $S_{\mu\nu}^{\rm M}$, in contrast to what is the case for the Abraham tensor (cf., for instance, Refs. \cite{nesterenko16a} or \cite{moller72}).

As should be clear, the  simplest tensor alternative to use in all existing experiments that we are aware of, is the Minkowski tensor. This tensor has gained considerable popularity also because it is so accessible to  a Lagrangian or Hamiltonian description for the field in matter. There exists by now a considerable literature on this topic; we will not attempt to cover that here, but restrict ourselves to a couple of our own contributions in that direction \cite{brevik17,breviklautrup70}. Especially in the work \cite{breviklautrup70} with Lautrup, we attempted to go into some detail.

\bigskip

\noindent {\it Relations to the Casimir effect.} There exists a delicate relationship  between macroscopic electrodynamics and the Casimir effect. The Casimir  effect, as known, is the attractive force between closely spaced dielectric bodies induced by  fluctuating electromagnetic fields (for an introduction cf., for instance, Ref.~\cite{bordag09}).  The point here is that the quantum field theory approach assumes the medium to be continuous; moreover  the classical electromagnetic boundary conditions are made use of.  Now, the continuum picture is known to be applicable  down to distances of the order of a few nanometers, but not for shorter distances at which a particle approach becomes mandatory.  In QFT for media, if  one uses  conventional time splitting regularization,  one will encounter  cutoff terms which in principle can be arbitrarily large,  reflecting  spatial separation distances that can be arbitrarily small. (There is an exception to this, namely the case of isorefractive media, but we will leave out this special case here.)  That is, the QFT picture becomes   applied to scales far beyond what is physically justifiable.  Faced with this dilemma one may wonder: is there after all a natural limitation of the cutoff parameters, related to physical quantities such as surface tension? This topic was recently commented on in Refs.~\cite{hoye17,parashar17}. Especially when considering a nonmagnetic spherical ball of radius $a$ \cite{hoye17} the result is worth attention: by equating the compressive cutoff term obtained from QFT to the compressive surface force density $2\sigma /a$ (with $\sigma$ the surface tension coefficient of conventional magnitude), one finds the time splitting time $\tau$ to be about $10^{-19}$ s.  This corresponds to a cutoff elementary distance of $\tau c \sim 1~${\AA}, which means atomic dimensions. So one may suggest: is there some kind of link between QFT and hydromechanics, limiting the values of the QFT cutoff parameters?

\section*{Acknowledgment}
This work is supported by the Research Council of Norway, Project No. 250346. Thanks go to  Prachi Parashar for comments on the manuscript. I have also benefited from a wide correspondence, with  Honggu Choi,  Nelson G. C. Astrath,  Mario Silveirinha, Masud Mansuripur,  Vladimir V. Nesterenko, and others.




\begin{thebibliography}{99}

\bibitem{abraham09} M. Abraham, Rend. Circ. Matem. Palermo {\bf 28}, 1 (1909).
\bibitem{abraham10} M. Abraham, Rend. Circ. Matem. Palermo {\bf 30}, 33 (1910).
\bibitem{minkowski10} H. Minkowski,  Math. Annaln. {\bf 68}, 472 (1910).
\bibitem{brevik79} I. Brevik, Phys. Rep. {\bf 52}, 133 (1979).
\bibitem{pfeifer07} R. N. Pfeifer, T. A. Nieminen, N. R. Heckenberg and H. Rubinsztein-Dunlop, Rev. Mod. Phys. {\bf 79}, 1197 (2007).
\bibitem{hinds09} E. A. Hinds and S. M. Barnett, Phys. Rev. Lett. {\bf 102}, 050403 (2009).
\bibitem{barnett10} S. M. Barnett and R. Loudon, Phil. Trans. R. Soc. A {\bf 368}, 927 (2010).
\bibitem{baxter10} C. Baxter and R. Loudon, J. Mod. Opt. {\bf 57}, 830 (2010).
\bibitem{barnett10a} S. M. Barnett, Phys. Rev. {\bf 104}, 070401 (2010).
\bibitem{mansuripur10} M. Mansuripur, Opt. Comm. {\bf 283}, 1997 (2010).
\bibitem{milonni10} P. W. Milonni and R. W. Boyd, Adv. Optics and Photonics {\bf 2}, 519 (2010).
\bibitem{kemp11} B. A. Kemp, J. Appl. Phys. {\bf 109}, 111101 (2011).
\bibitem{rikken11} G. L. J. A. Rikken and B. A. van Tiggelen, Phys. Rev. Lett. {\bf 107}, 170401 (2011).
\bibitem{griffiths12} D. J. Griffiths, Am. J. Phys. {\bf 80}, 7 (2012).
\bibitem{testa13} M. Testa, Ann. Phys. (N.Y.) {\bf 336}, 1 (2013).
\bibitem{mansuripur13} M. Mansuripur, A. R. Zakharian and E. M. Wright, Phys. Rev. A {\bf 88}, 023826 (2013).
\bibitem{aanensen13} N. S. Aanensen, S. {\AA.} Ellingsen and I. Brevik, Phys. Scripta {\bf 87}, 055402 (2013).
\bibitem{astrath14} N. G. C. Astrath, L. C. Malacarne, M. L. Baesso, G. V. B. Lukasievicz and S. E. Bialkowski, Nature Commun. {\bf 5}, 4363 (2014) [DOI: 10.1038/ncomms5363].
\bibitem{leonhardt14} U. Leonhardt, Phys. Rev. A {\bf 90}, 033801(2014).
\bibitem{brevikhoye14} I. Brevik, Transactions of the Royal Norwegian Society of Sciences and Letters (2014)(3), p. 83; arXiv:1310.3684 [quant-ph].
\bibitem{barnett15} S. M. Barnett and R. Loudon, New. J. Phys. {\bf 17}, 063027 (2015).
\bibitem{zhang15}L. Zhang, W. She, N. Peng and U. Leonhardt, New J. Phys. {\bf 17}, 053035 (2015).
\bibitem{kemp15} B. A. Kemp, Prog. Opt. {\bf 60}, 437 (2015).
\bibitem{sheppard16} C. J. Sheppard and B. A. Kemp, Phys. Rev. A {\bf 93}, 053832 (2016).
\bibitem{wang16}S. Wang, J. Ng, M. Xiao and C. T. Chan, Scient. Adv. {\bf 2}, e1501485 (2016).
\bibitem{nesterenko16} V. V. Nesterenko and A. V. Nesterenko, J. Math. Phys. {\bf 57}, 032901 (2016).
\bibitem{nesterenko16a} V. V. Nesterenko and A. V. Nesterenko, J. Math. Phys. {\bf 57}, 092902 (2016).
\bibitem{brevik17}I. Brevik, Ann. Phys. (N.Y.) {\bf 377}, 10 (2017).
\bibitem{choi17} H. Choi, M. Park, D. S. Elliot, and K. Oh, Phys. Rev. A {\bf 95}, 053817 (2017).
\bibitem{silveirinha17} M. G. Silveirinha, Phys. Rev. A {\bf 96}, 033831 (2017).
\bibitem{partanen17}M. Partanen, T. H{\"a}yrynen, J. Oksanen, and J. Tulkki, Phys. Rev. A {\bf 95}, 063850 (2017).
\bibitem{partanen17a}M. Partanen and J. Tulkki, Phys. Rev. A {\bf 96}, 063834 (2017).
\bibitem{medina17}R. Medina and J. Stephany, arXiv:1703.02109 [physics.class-ph].
\bibitem{landau84} L. D. Landau and E. M. Lifshitz, {\it Electrodynamics of Continuous Media}, 2nd ed. (Butterworth-Heinemann, Oxford, 1984).
\bibitem{lai89} H. M. Lai, P. T. Leung, K. L. Poon, and K. Young, J. Opt. Soc. Am. B {\bf 6}, 2430 (1989).
\bibitem{zimmerli99} G. A. Zimmerli, R. A. Wilkinson, R. A. Ferrell, and M. R. Moldover, Phys. Rev. E {\bf 59}, 5802 (1999).
\bibitem{ellingsen11}S. {\AA.} Ellingsen and I. Brevik, Phys. Fluids {\bf 23}. 096101 (2011).
\bibitem{hakim62} S. S. Hakim and J. B. Higham, Proc. Phys. Soc. London {\bf 80}, 190 (1962).
\bibitem{ashkin73} A. Ashkin and J. M. Dziedzic, Phys. Rev. Lett. {\bf 30}, 19 (1973).
\bibitem{casner03} A. Casner and J. P. Delville, Phys. Rev. Lett. {\bf 90}, 144503 (2003).
\bibitem{wunenburger11} R. Wunenburger, B. Issenmann, E. Brasselet, C. Loussert, V. Houstane and J. P. Delville, J. Fluid Mech. {\bf 666}, 273 (2011).
\bibitem{hallanger05}A. Hallanger, I. Brevik, S. Haaland and R. Sollie, Phys. Rev. E {\bf 71}, 056601 (2005).
\bibitem{birkeland08} O. J. Birkeland and I. Brevik, Phys. Rev. E {\bf 78}, 066314 (2008).
\bibitem{capeloto16} O. A. Capeloto, V. S. Zanuto, L. C. Malacarne, M. L. Baesso, G. V. B. Lukasievicz, S. E. Bialkowski, and N. G. C. Astrath, Scient. Rep. {\bf 6}, 20515; doi:10.11038/srep20515(2016).
\bibitem{zhang88} J. Z. Zhang and R. K. Chang, Opt. Lett. {\bf 13}, 916 (1988).
\bibitem{brevik99} I. Brevik and R. Kluge, J. Opt. Soc. B {\bf 16}, 976 (1999).
\bibitem{ellingsen13} S. {\AA.} Ellingsen, J. Opt. Soc. B {\bf 30}, 1694 (2013).
\bibitem{kundu17} A. Kundu, R. Rani, and K. S. Hazra, Scient. Reports {\bf 7}, 42538 (2017); doi:10.1038/srep42538.
\bibitem{wang08}X. Wang, Y. P. Cheng, and D. D. Nolte, Opt. Express {\bf 16}, 22105 (2008).
\bibitem{jones54} R. V. Jones and J. C. Richards, Proc. R. Soc. A {\bf 221}, 480 (1954).
\bibitem{jones78} R. V. Jones and B. Leslie, Proc. R. Soc. A {\bf 360}, 347 (1978).
\bibitem{jones88}R. V. Jones, {\it Instruments and Experiences} (John Wiley \& Sons, New York, 1988).
\bibitem{gibson80} A. F. Gibson, M. F. Kimmitt, A. O. Koohian, D. E. Evans, and G. F. D. Levy, Proc. Roy. Soc. London Ser. A {\bf 370}, 303 (1980).
\bibitem{brevik86} I. Brevik, Phys. Rev. B {\bf 33}, 1058 (1986).
\bibitem{campbell05} G. K. Campbell, A. E. Leanhardt, J. Mun, M. Boyd, E. W. Streed, W. Ketterle and D. E. Pritchard, Phys. Rev. Lett. {\bf 94}, 170403 (2005).
\bibitem{brevik10} I. Brevik and S. {\AA}. Ellingsen, Phys. Rev. A {\bf 81}, 063830 (2010).
\bibitem{walker75a} G. B. Walker, D. G. Lahoz and G. Walker, Can. J. Phys. {\bf 53}, 2577 (1975).
\bibitem{walker75b} G. B. Walker and D. G. Lahoz, Nature (London) {\bf 253}, 339 (1975).
\bibitem{balasz53}N. L. Balazs, Phys. Rev. {\bf 91}, 408 (1953).
\bibitem{dodin12} I. Y. Dodin and N. J. Fisch, Phys. Rev. A {\bf 86}, 053834 (2012).
\bibitem{crenshaw14} M. E. Crenschaw, J. Math. Phys. {\bf 55}, 042901 (2014).
\bibitem{crenshaw16} M. E. Crenschaw, arXiv:1604.01801 [physics.optics].
\bibitem{moller72} C. M{\o}ller, {\it The Theory of Relativity}, 2nd ed. (Clarendon Press, Oxford, 1972).
\bibitem{breviklautrup70} I. Brevik and B. Lautrup, Mat. Fys. Medd. K. Dan. Vid. Selsk. {\bf 38}(1), 1 (1970). Available at the library  www.sdu.dk/Bibliotek/
\bibitem{bordag09} M. Bordag, G. I. Klimchitskaya, U. Mohideen, and V. M. Mostepanenko, {\it Advances in the Casimir Effect} (Oxford University, Oxford, 2009).
\bibitem{hoye17} J. S. H{\o}ye and I. Brevik, Phys. Rev. A {\bf 95}, 052127 (2017).
\bibitem{parashar17} P. Parashar, K. A. Milton, K. V. Shajesh, and I. Brevik, Phys. Rev. D {\bf 96}, 085010 (2017).


\vspace{2cm}









\end{thebibliography}
\end{document}